\documentclass[aps,prb,twocolumn,superscriptaddress,floatfix,longbibliography]{revtex4-2}

\usepackage{amsmath,amssymb} 
\usepackage{bm} 
\usepackage{graphicx} 
\usepackage{comment} 
\usepackage{textcomp} 
\usepackage[acronym]{glossaries} 
\usepackage{float}
\usepackage{upgreek}

\usepackage{siunitx}

\newacronym{qcl}{QCL}{quantum cascade laser}
\newacronym{fp}{FP}{Fabry-Perot}
\newacronym{rt}{RT}{racetrack}
\newacronym{wg}{WG}{waveguide}

\usepackage{enumitem}
\setlist{noitemsep,leftmargin=*,topsep=0pt,parsep=0pt}

\usepackage{xcolor} 
\definecolor{lightgray}{gray}{0.6}
\definecolor{medgray}{gray}{0.4}

\usepackage{hyperref}
\hypersetup{
colorlinks=true,
urlcolor= blue,
citecolor=blue,
linkcolor= blue,
}

\newif\ifptitle
\newif\ifpnumber
\newcounter{para}

\ptitletrue  
\pnumbertrue  

\newcommand{\mytitle}{Semiconductor ring laser frequency combs with active directional couplers}

\begin{document}

\title{\mytitle}

\author{Dmitry Kazakov}
\email[]{kazakov@seas.harvard.edu}
\affiliation{Harvard John A. Paulson School of Engineering and Applied Sciences, Harvard University, Cambridge, MA 02138, USA}
\author{Theodore P. Letsou}
\affiliation{Harvard John A. Paulson School of Engineering and Applied Sciences, Harvard University, Cambridge, MA 02138, USA}
\affiliation{Department of Electrical Engineering and Computer Science, Massachusetts Institute of Technology, Cambridge, MA 02142, USA}

\author{Maximilian Beiser}
\affiliation{Institute of Solid State Electronics, TU Wien, 1040 Vienna, Austria}

\author{Yiyang Zhi}
\affiliation{Harvard John A. Paulson School of Engineering and Applied Sciences, Harvard University, Cambridge, MA 02138, USA}
\affiliation{Case Western Reserve University, Cleveland, OH 44106, USA}




\author{Nikola Opa$\mathrm{\check{c}}$ak}
\affiliation{Harvard John A. Paulson School of Engineering and Applied Sciences, Harvard University, Cambridge, MA 02138, USA}
\affiliation{Institute of Solid State Electronics, TU Wien, 1040 Vienna, Austria}


\author{Marco Piccardo}
\affiliation{Harvard John A. Paulson School of Engineering and Applied Sciences, Harvard University, Cambridge, MA 02138, USA}
\affiliation{Center for Nano Science and Technology, Fondazione Istituto Italiano di Tecnologia, Milano, Italy}

\author{Benedikt Schwarz}
\affiliation{Harvard John A. Paulson School of Engineering and Applied Sciences, Harvard University, Cambridge, MA 02138, USA}
\affiliation{Institute of Solid State Electronics, TU Wien, 1040 Vienna, Austria}

\author{Federico Capasso}
\email[]{capasso@seas.harvard.edu}
\affiliation{Harvard John A. Paulson School of Engineering and Applied Sciences, Harvard University, Cambridge, MA 02138, USA}


\begin{abstract}
Rapid development of Fabry-Perot quantum cascade laser frequency combs has converted them from laboratory devices to key components of next-generation fast molecular spectrometers. Recently, free-running ring quantum cascade lasers allowed generation of new frequency comb states induced by phase turbulence. In absence of efficient light outcoupling, ring quantum cascade lasers are not suited for applications as they are limited in their power output to microwatt levels. 
Here we demonstrate electrically pumped ring quantum cascade lasers with integrated active directional couplers. These devices generate self-starting frequency combs and have output power above ten milliwatts at room temperature. We study the transmission of the ring-waveguide resonator system below the lasing threshold, which reveals the ability to individually control the mode indices in the coupled resonators, their quality factors, and the coupling coefficient. When the ring resonator is pumped above the lasing threshold, the intracavity unidirectional single-mode field parametrically amplifies an externally injected signal tuned into one of the ring resonances, generating an idler sideband via four-wave mixing. The ability to inject external optical signals into integrated laser cavities brings into reach coherent control of frequency comb states in ring semiconductor lasers. Furthermore, tunable coupled active resonators pumped below the lasing threshold enable a versatile platform for the studies of resonant electromagnetic effects, ranging from strong coupling to parity-time symmetry breaking.

\end{abstract}

\maketitle

\begin{figure*}[t]
    \includegraphics[clip=true,width=\textwidth]{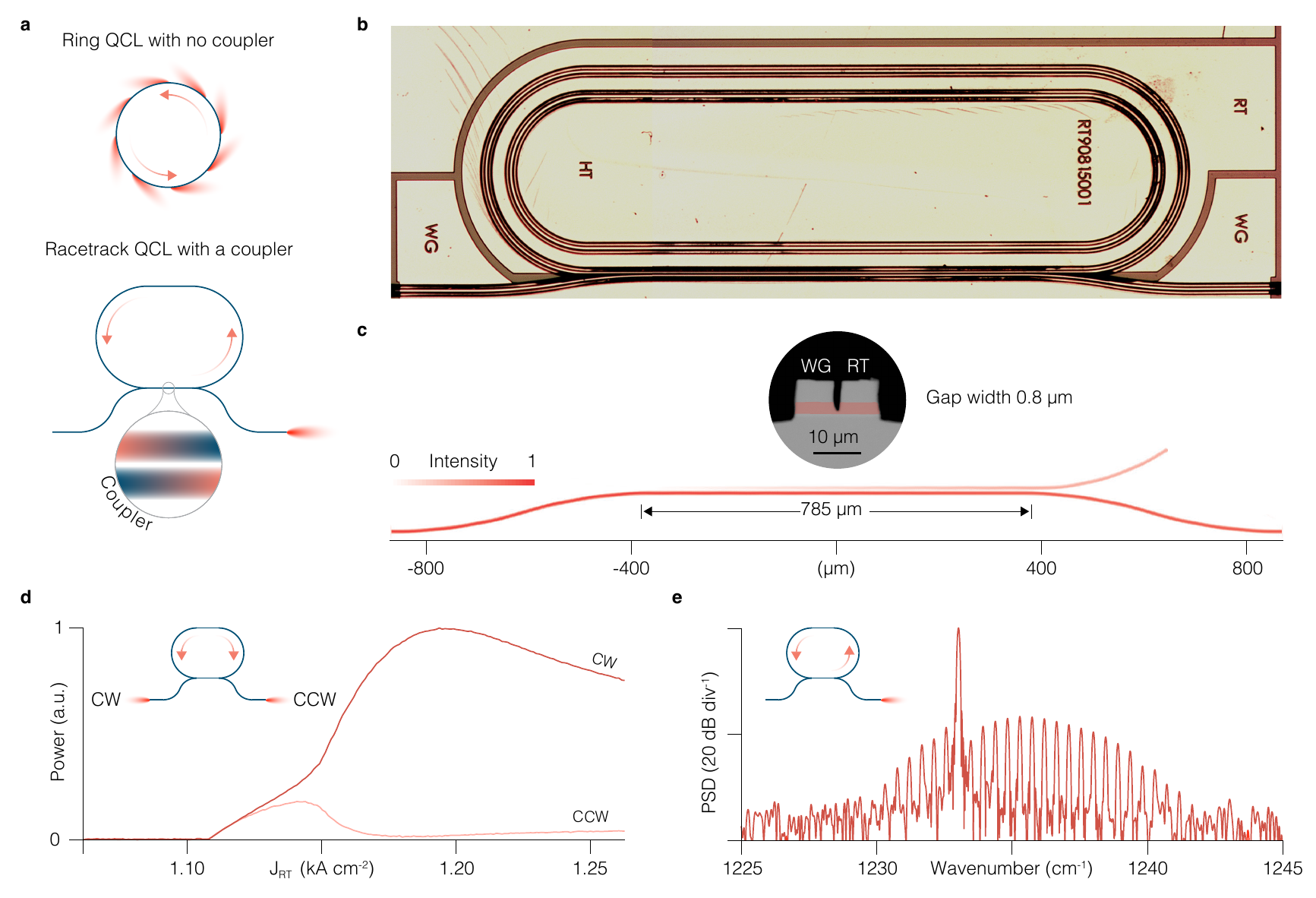}
    \caption{\textbf{Racetrack quantum cascade lasers with active directional couplers.} \textbf{a}, Light outcoupling from a ring cavity versus a racetrack cavity with an integrated directional coupler. \textbf{b}, Optical microscope image of the racetrack QCL with an integrated active directional coupler. Contact sections for separate biasing are denoted with \acrshort{rt}, for the racetrack, \acrshort{wg}, for the waveguide coupler and HT, for the integrated heater. CW and CCW denote the outcoupling ports for the clockwise and counterclockwise fields. \textbf{c}, Simulated intensity distribution in the coupling region. The inset shows an optical microscope image of the cross-section of the coupling region. The waveguide cores filled with laser active medium are colored in red. \textbf{d}, Experimental intensities as function of the injected current collected simultaneously from both WG ports on two external detectors, showing the regimes of bidirectional and unidirectional lasing. \textbf{e}, Experimental spectrum of the self-starting frequency comb in a racetrack QCL, when it operates in a unidirectional regime. The spectrum has a bell-shaped envelope with a pronounced primary mode, characteristic of ring \acrshort{qcl}s.}
     \label{fig:racetrack_device}
\end{figure*}


\begin{figure*}[t]
    \includegraphics[clip=true,width=\textwidth]{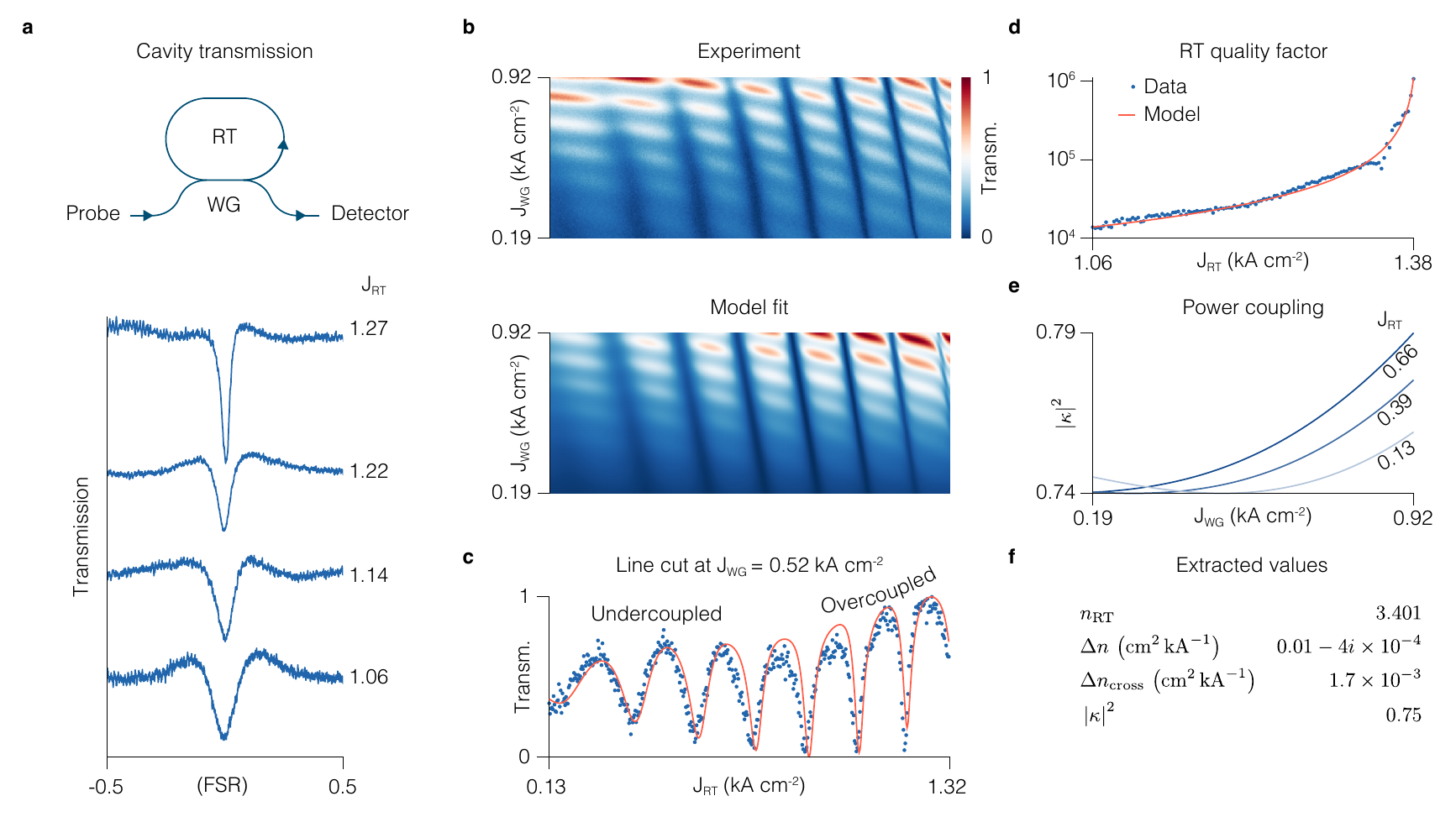}
    \caption{\textbf{Transmission characteristics of the racetrack-waveguide system.} \textbf{a}, Experimental transmission of the RT coupled to the WG as the wavelength of the probe laser is swept across the RT resonance. The resonance narrows for increasing RT current density. \textbf{b}, Transmitted intensity of the probe signal at $1222$ cm$^{-1}$ as function of current density of the racetrack QCL (\acrshort{rt}) and the directional waveguide coupler (\acrshort{wg}) and the corresponding least squares model fit. The experiment shows the ability to control complex mode indices in the coupled resonators. \textbf{c}, Experimental transmission (dots) and model fit (solid line) as function of the RT current density for the fixed WG current density of 0.52 kA cm$^{-2}$. \textbf{d}, Extracted (dots) and modeled (solid line) \acrshort{rt} resonance quality factor as function of the \acrshort{rt} current.  \textbf{e}, Power coupling coefficient as function of the WG current for three different values of the RT current. \textbf{f}, Values extracted from the fit. $n_{\mathrm{RT}}$, base mode index of the RT. $\Delta n$, complex index change per unit of current density. $\Delta n_{\mathrm{cross}}$, index change due to thermal crosstalk between WG and RT. $\kappa$, power coupling coefficient.}
     \label{fig:subthreshold_cavity_spectroscopy}
\end{figure*}

\begin{figure*}[t]
    \includegraphics[clip=true,width=\textwidth]{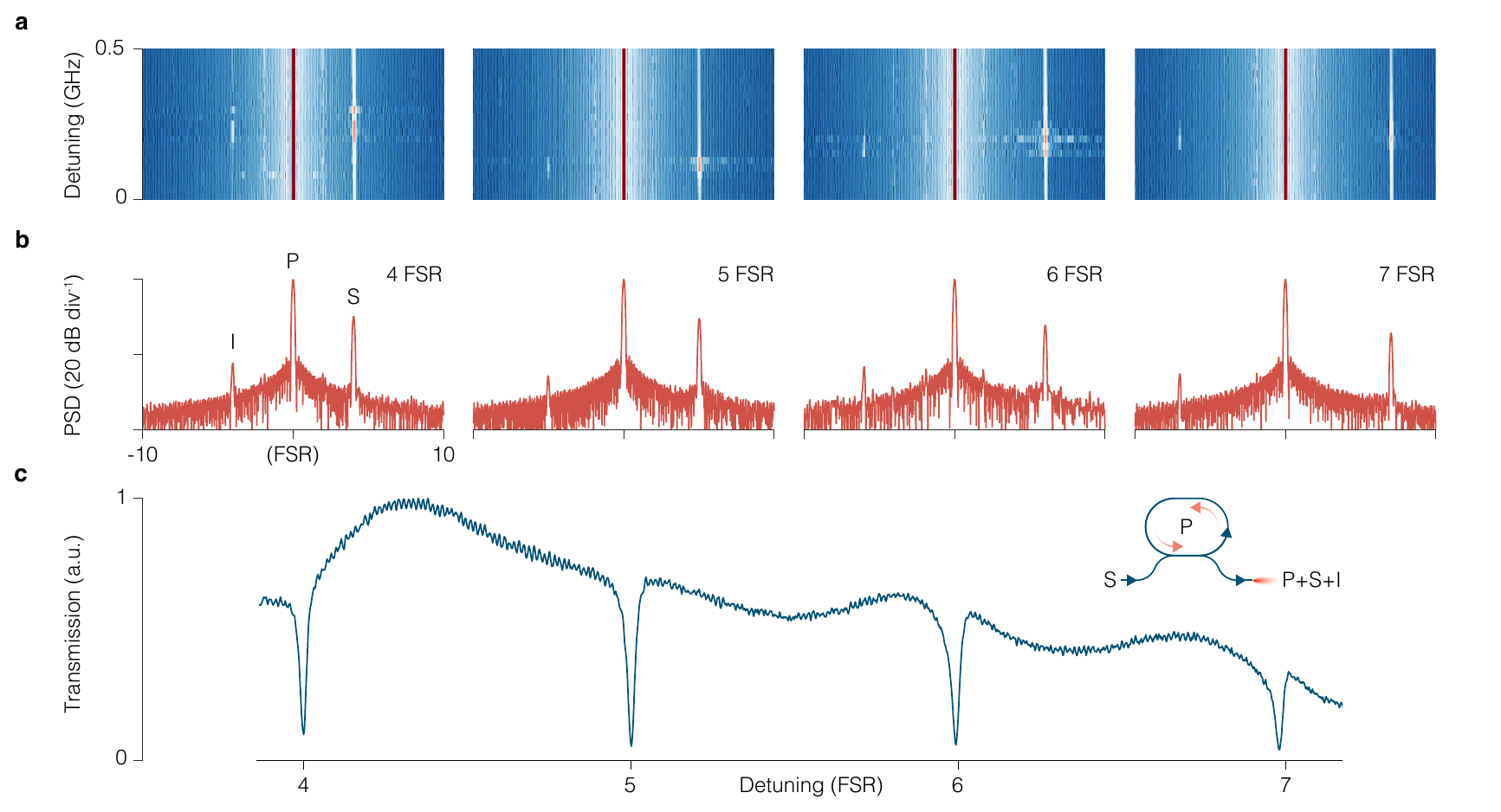}
    \caption{\textbf{Parametric sideband generation by four-wave mixing.} \textbf{a}, Experimental spectrograms of the racetrack QCL above the threshold under external optical injection as the detuning of the signal is swept in the vicinity of the racetrack resonance, shown for four subsequent resonances, 4, 5, 6 and 7 FSRs away from the racetrack lasing frequency. \textbf{b}, Optical spectra for the detuning of the signal wave (S) when the idler wave (I) is the strongest. P, the pump wave generated by the racetrack above its lasing threshold. \textbf{c}, Experimental transmission, below the lasing threshold of the racetrack, showing four resonances that we inject into to generate idler sidebands.}
     \label{fig:FWM}
\end{figure*}


Compact chip-scale frequency comb generators hold potential to revolutionise fiber optic communications, optical ranging, and trace gas spectroscopy~\cite{Chang2022IntegratedTechnologies,Shams-Ansari2022Thin-filmSpectroscopy,Villares2014Dual-combCombs,Riemensberger2020MassivelyMicrocomb,Marin-Palomo2017Microresonator-basedCommunications,Dutt2018On-chipSpectroscopy}. Cumulatively, they cover the spectral range from the visible down to THz and come in a variety of form factors and material platforms, yet they can be broadly categorized into two distinct groups, based on the physical mechanism behind comb formation: actively or passively mode-locked semiconductor lasers and integrated nonlinear microresonator combs. Quantum cascade laser (\acrshort{qcl}) frequency combs lie at the intersection of the two categories: the ultrashort gain recovery time prevents traditional passive mode-locking in \acrshort{qcl}s, yet they can be actively modulated to produce mode-locked pulses of picosecond duration straight out of the cavity and reach the subpicosecond regime with external pulse compression~\cite{Hillbrand2020Mode-lockedLaser, Taschler2021FemtosecondLaser}. Even in the absence of modulation or any additional intracavity elements, free-running Fabry-Perot (FP) \acrshort{qcl}s generate highly phase-coherent, broadband frequency-modulated combs thanks to the combination of unclamped  incoherent gain due to a population inversion grating caused by strong spatial hole burning, and parametric gain due to the interplay between cavity dispersion and giant optical nonlinearities of the gain medium itself~\cite{Mansuripur2016Single-modeOscillator,Kazakov2017Self-startingLaser,Burghoff2020UnravelingTheory,Forrer2021Self-startingLasers,Piccardo2022LaserApplications,Opacak2021FrequencyNonlinearity}. 

Recently, ring \acrshort{qcl} combs were demonstrated in the mid-infrared and THz~\cite{Piccardo2020FrequencyTurbulence,Meng2020Mid-infraredLaser,Jaidl2022SiliconComb}. In ring \acrshort{qcl}s operating in unidirectional regime the population grating is absent and the incoherent gain is clamped above lasing threshold. As a result, it is only the  parametric gain that is responsible for multimode emission via phase turbulence and eventual comb formation~\cite{Piccardo2020FrequencyTurbulence,Kazakov2021Defect-engineeredCombs}. The parametric gain originates from fluctuations of both the amplitude and phase of the intracavity field, which are reinforced by the modulation of both the real and imaginary part of the optical susceptibility of the medium, quantified by a nonzero linewidth enhancement factor~\cite{Opacak2021SpectrallyComb}. 
Ring \acrshort{qcl} combs thus fall in between traditional electrically pumped semiconductor laser frequency combs and Kerr combs, where the instability gain is provided by self-phase modulation of the external optical pump field in presence of the intensity-dependent refractive index~\cite{Kippenberg2011Microresonator-basedCombs,Godey2014StabilityRegimes}. 

The relation between the active ring resonator \acrshort{qcl} combs and passive resonator Kerr combs was formalized in the recently developed generalized theory of frequency combs in active and passive media~\cite{Columbo2021UnifyingLasers}. The theory unified the two platforms and provided a scheme for coherent control of the comb states in ring \acrshort{qcl}s by means of injection of an external optical signal. Experimental verification of the theory requires efficient coupling of external radiation. So far experimental demonstrations of ring \acrshort{qcl} combs were done in circular cavities where the light outcoupling was due to bending losses through the curved waveguide sidewall (Fig.~\ref{fig:racetrack_device}\textbf{a}), thus leaving both optical power extraction and external injection efficiency at a fraction of a percent --- prohibitively small to make these devices attractive for applications or to allow coherent state control via optical injection.  


The ability to efficiently extract power from ring \acrshort{qcl}s and to coherently control their state requires the integration of a coupling waveguide adjacent to the ring resonator. (Fig.~\ref{fig:racetrack_device}\textbf{a}). There exists a number of ways to implement such an integrated waveguide. The basic paradigm of photonic integration, regardless of the employed scheme --- monolithic, heterogeneous or hybrid --- relies on delegating the generation, modulation, and detection of light to active components made of III-V semiconductor compounds while utilizing passive  low-loss dielectrics for on-chip light guiding~\cite{Huang2020EpitaxialFeedback,Wang2022MonolithicWaveguides,Bhardwaj2020AModulator}. Active-passive integration schemes demand considerable fabrication effort, as they rely on either wafer bonding, where achieving alignment of the passive and active components is the main challenge, or multiple step epitaxial regrowth, where issues may arise due to lattice mismatch. On the other hand, making waveguides from the very same active medium that is used for light generation has proven successful for semiconductor ring lasers~\cite{Gelens2009ExploringExperiment,Nshii2010ALaser,Kacmoli2022UnidirectionalLasers}. Such purely active integration scheme is attractive due to its relative ease of implementation, since the active waveguides are defined on the same lithographic and etching step as the laser cavities. As we show in what follows, active waveguides naturally allow for fine tuning of the mode indices and losses, enabling control over resonance detuning, quality factor, and power coupling coefficient. 


We designed and fabricated ring \acrshort{qcl}s shaped as racetracks with optical coupling ports which are implemented as evanescent wave directional couplers, where the input waveguide is separated by a narrow air gap from the straight section of the racetrack (Fig.~\ref{fig:racetrack_device}\textbf{b}). The coupling ratio is set at the design stage by the length of the interaction region and by the width of the air gap (Fig.~\ref{fig:racetrack_device}\textbf{c}). For a gap width of $0.8$ \si{\micro\meter}, attainable using optical lithography, we simulate a power coupling ratio of -10 dB (Supplementary Information). The fabricated devices feature slightly underetched gaps, which translate into a coupling coefficient higher than its design value (Fig.~\ref{fig:racetrack_device}\textbf{c}). Both the waveguide coupler (WG) and the racetrack (RT) contain the \acrshort{qcl} gain medium and have separate contacts that enable their independent driving. Individual control over injected electrical currents allows for fine tuning of the coupling strength, as asymmetric electrical driving results in a mismatch of mode indices inside the WG and RT (Supplementary Information). Furthermore, as the WG contains active media, it acts as a semiconductor optical amplifier for the light coupled in and out of the RT when pumped above transparency~\cite{Menzel2011QuantumK}. 


RT QCLs with directional couplers are superior to circular ring QCLs without coupling ports, in that they enable the extraction of optical power at levels that are suitable for applications. In circular ring QCLs without coupling ports only a small fraction of light circulating inside the cavity is outcoupled, which in previous studies resulted in an optical output of at most half a milliwatt at room temperature~\cite{Meng2020Mid-infraredLaser,Piccardo2020FrequencyTurbulence}. On the other hand, the RT QCLs presented here feature output power levels above 10 mW at room temperature, comparable to FP QCLs of similar length and identical waveguide width fabricated on the same wafer (Supplementary Information). 

RT \acrshort{qcl}s operate in a unidirectional regime above the threshold, as we show by simultaneously detecting the light intensity coming out of both ends of the \acrshort{wg} coupler while increasing the \acrshort{rt} pump current (Fig.~\ref{fig:racetrack_device}\textbf{d}). Right above the threshold clockwise (CW) and counterclockwise (CCW) waves grow at same rate, up until the point of symmetry breaking where CW wave intensity starts growing at a higher rate at the expense of the decreasing intensity of the CCW wave, ultimately leading to lasing in CW direction only at higher pumping level (Fig.~\ref{fig:racetrack_device}\textbf{d}). RT QCLs with couplers can thus attain a unidirectional regime despite potentially higher backscattering, than circular rings without couplers, due to the index change within the coupler section and reflections off the uncoated end facets of the \acrshort{wg}. Unidirectional operation is essential to eliminate spatial hole burning and to facilitate comb generation due to the parametric gain only, as the incoherent gain remains clamped above the symmetry breaking point. Once the parametric gain is high enough, phase turbulence leads to an instability resulting in the formation of a frequency comb with a characteristic bell-shaped spectral envelope (Fig.~\ref{fig:racetrack_device}\textbf{e})~\cite{Meng2021DissipativeLasers}.

To demonstrate the ability to couple light from an external laser into the ring cavity, we perform spectroscopy of the RT-WG system while keeping the RT below its lasing threshold (Fig.~\ref{fig:subthreshold_cavity_spectroscopy}\textbf{a}). For the injection, we use a FP \acrshort{qcl}, operating in a single-mode regime, that is focused onto a facet of the \acrshort{wg} with an aspheric antireflection coated lens (NA $=0.56$). 
Light transmitted through the WG is collected at the opposite facet with an identical lens and focused onto an HgCdTe infrared photodetector. To eliminate detection of the amplified spontaneous emission from the WG and to increase the signal-to-noise ratio we use lock-in detection by chopping the probe beam at a rate of 320 Hz. 

First we sweep the wavelength of the probe laser across one \acrshort{rt} resonance while keeping the \acrshort{wg} bias constant. As an evidence of coupling to the \acrshort{rt}, the transmission as a function of probe detuning shows a dip, whose linewidth decreases as we increase the electrical pumping of \acrshort{rt} --- the resonator quality factor grows as the intersubband absorption decreases (Fig.~\ref{fig:subthreshold_cavity_spectroscopy}\textbf{a}). The quality factor of the \acrshort{rt} can be continuously tuned over several orders of magnitude (Fig.~\ref{fig:subthreshold_cavity_spectroscopy}\textbf{d}). Next we fix the wavelength of the probe \acrshort{fp} \acrshort{qcl} at 1222 cm$^{-1}$ and sweep the drive currents of both \acrshort{wg} and \acrshort{rt}. Increasing the \acrshort{rt} current leads to several effects in the recorded transmission. First, the thermal change of the mode index effectively sweeps the \acrshort{rt} cavity mode resonances with respect to the fixed probe laser frequency (Fig.~\ref{fig:subthreshold_cavity_spectroscopy}\textbf{b}). Second increasing pumping decreases the intersubband absorption in the \acrshort{rt} active medium, which leads to the narrowing of the resonances (Fig.~\ref{fig:subthreshold_cavity_spectroscopy}\textbf{b}, \textbf{c}). 
Third, the mode index mismatch between the coupled waveguides caused by the asymmetric driving of the \acrshort{wg} and \acrshort{rt} leads to the change of the coupling strength between the \acrshort{wg} and \acrshort{rt} (Fig.~\ref{fig:subthreshold_cavity_spectroscopy}\textbf{e}). The last two effects combined allows one to traverse different coupling regimes of the system, from undercoupling to critical coupling to overcoupling, depending on the ratio between the gain inside the RT and the coupling coefficient (Fig.~\ref{fig:subthreshold_cavity_spectroscopy}\textbf{c}). 

The resonant structure of the transmission appears as well when sweeping the current of the \acrshort{wg}. Since the reflectivities of the uncoated facets of the \acrshort{wg} are $0.3$, when biased at or below transparency it is effectively a low-finesse FP resonator coupled to a high-finesse \acrshort{rt} resonator. At low \acrshort{wg} bias its resonant structure is washed out due to high intersubband absorption. Increasing the \acrshort{wg}  bias leads to a decrease in intersubband absorption as the medium approaches inversion. At high bias of the \acrshort{wg} and \acrshort{rt}, both resonators attain a high finesse and the resonance profiles show avoided mode crossing --- a signature of strongly coupled resonators (Fig.~\ref{fig:subthreshold_cavity_spectroscopy}\textbf{b}).  

The transmission characteristic of the \acrshort{rt}-\acrshort{wg} system agrees remarkably well with a simple transfer matrix model that we adapt from Ref.~\cite{Yariv2006TransmissionSystem} to account for the complex nature of the mode indices and for the thermal crosstalk between the waveguides (Fig.~\ref{fig:subthreshold_cavity_spectroscopy}\textbf{b}, \textbf{c} and Supplementary Information). The model captures all features of the experimental transmission, and allows us to extract relevant device parameters, such as base complex mode indices, their change per unit of current density, power coupling coefficient between the \acrshort{rt} and the \acrshort{wg} and the amount of thermal crosstalk (Fig.~\ref{fig:subthreshold_cavity_spectroscopy}\textbf{f}). The excellent agreement between the model and experimental data shows that the presented devices are well-behaved under external probing, hence are ideally suited for future studies of frequency comb dynamics in an externally driven scheme~\cite{Columbo2021UnifyingLasers}. 


As a step towards external coherent control of the laser state, we bring the \acrshort{rt} above the lasing threshold such that it generates a unidirectional strong field that serves as a pump for the parametric gain experienced by the externally injected weak probe signal. We sweep the wavelength of the injected signal through four subsequent resonances of the \acrshort{rt} resonator, on the blue side of the \acrshort{rt} pump field (Fig.~\ref{fig:FWM}\textbf{c}). When tuning into each of the four resonances we observe an appearance of the idler sideband symmetrically on the red side of the pump --- a signature of parametric amplification via four-wave mixing provided by the coherent interaction of the pump and the signal waves (Fig.~\ref{fig:FWM}\textbf{a}, \textbf{b})~\cite{Friedli2013Four-waveAmplifier}. With higher power of the probe laser it should be possible to achieve further mode proliferation and generation of harmonic frequency combs with an on-demand spacing~\cite{Piccardo2018WidelyLaser}. Bringing the probe laser frequency in the vicinity of the pump field should enable generation of bright and dark cavity solitons and Turing rolls, allowing for deterministic coherent control of the laser state~\cite{Columbo2021UnifyingLasers}.  


Additionally, the demonstrated devices, thanks to the ability of controlling coupling, resonance detuning, and quality factors below the lasing threshold, may serve as a versatile testbed for the various regimes of photonic resonances and allow continuous tuning between weak coupling, Fano resonances, strong coupling, PT-symmetry, and PT-symmetry breaking  --- all in one, all-electrically controlled integrated electro-optic device~\cite{Limonov2017FanoPhotonics}. The shown design, fabrication, and characterization methods are directly applicable to the implementation of a system of two evanescently coupled ring resonators, one- and two-dimensional ring resonator arrays, where each resonator can be individually addressed via electrical pumping. When operated above the threshold, these systems may provide a rich nonlinear optical playground for the studies of coherent interaction of the frequency comb states, their synchronization, and topological effects~\cite{Mittal2021TopologicalSolitons}.

    

\section*{Acknowledgements}
T. P. Letsou would like to thank the support of the Department of Defense (DoD) through the National Defense Science and Engineering Graduate (NDSEG) Fellowship Program. N. Opa$\mathrm{\check{c}}$ak and B. Schwarz are supported by the European Research Council (853014).

\bibliography{racetrack_references}
\clearpage

\end{document}


\title{\mytitle}

\author{Dmitry Kazakov}
\email[]{kazakov@seas.harvard.edu}
\affiliation{Harvard John A. Paulson School of Engineering and Applied Sciences, Harvard University, Cambridge, MA 02138, USA}
\author{Theodore P. Letsou}
\affiliation{Harvard John A. Paulson School of Engineering and Applied Sciences, Harvard University, Cambridge, MA 02138, USA}
\affiliation{Department of Electrical Engineering and Computer Science, Massachusetts Institute of Technology, Cambridge, MA 02142, USA}

\author{Maximilian Beiser}
\affiliation{Institute of Solid State Electronics, TU Wien, 1040 Vienna, Austria}

\author{Yiyang Zhi}
\affiliation{Harvard John A. Paulson School of Engineering and Applied Sciences, Harvard University, Cambridge, MA 02138, USA}
\affiliation{Case Western Reserve University, Cleveland, OH 44106, USA}




\author{Nikola Opa$\mathrm{\check{c}}$ak}
\affiliation{Harvard John A. Paulson School of Engineering and Applied Sciences, Harvard University, Cambridge, MA 02138, USA}
\affiliation{Institute of Solid State Electronics, TU Wien, 1040 Vienna, Austria}


\author{Marco Piccardo}
\affiliation{Harvard John A. Paulson School of Engineering and Applied Sciences, Harvard University, Cambridge, MA 02138, USA}
\affiliation{Center for Nano Science and Technology, Fondazione Istituto Italiano di Tecnologia, Milano, Italy}

\author{Benedikt Schwarz}
\affiliation{Harvard John A. Paulson School of Engineering and Applied Sciences, Harvard University, Cambridge, MA 02138, USA}
\affiliation{Institute of Solid State Electronics, TU Wien, 1040 Vienna, Austria}

\author{Federico Capasso}
\email[]{capasso@seas.harvard.edu}
\affiliation{Harvard John A. Paulson School of Engineering and Applied Sciences, Harvard University, Cambridge, MA 02138, USA}


\maketitle
\section{Directional coupler design}

\begin{figure}[t!]
    \centering
    \includegraphics[width=1.0\textwidth,center]{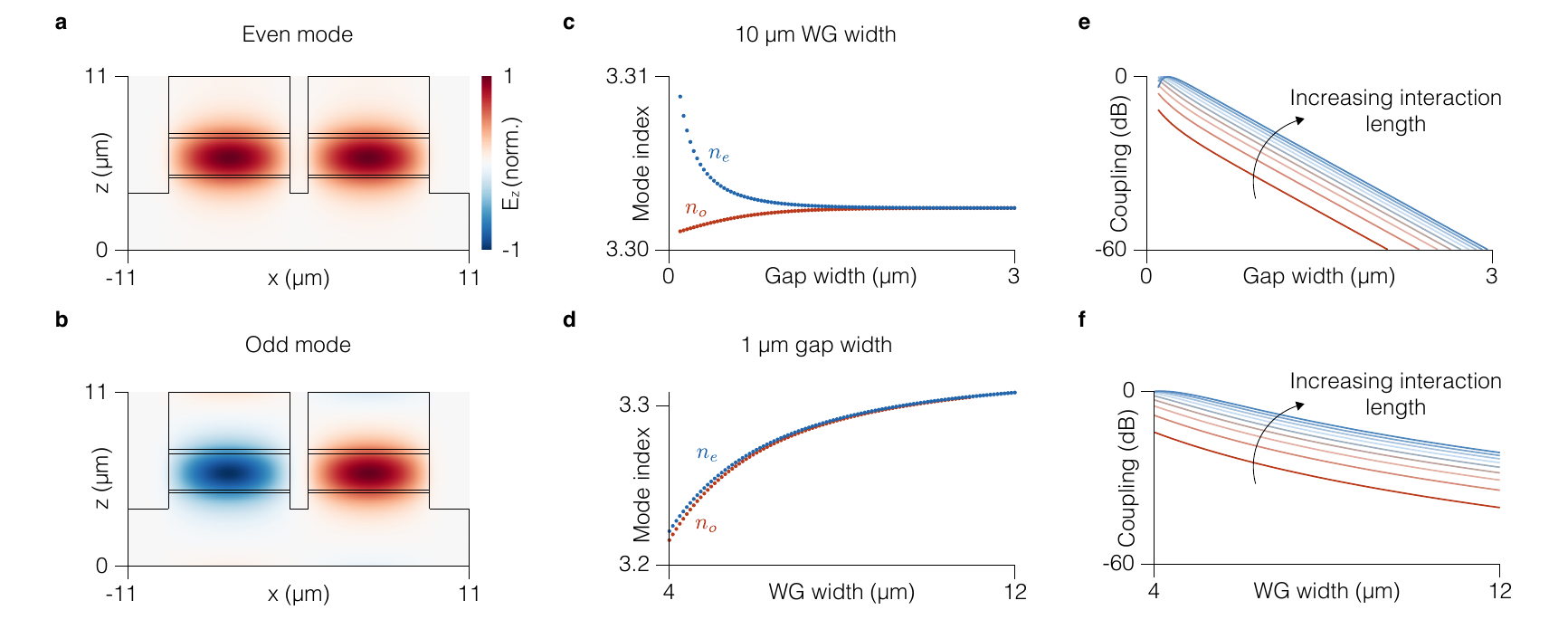}
    \centering
    \caption{\textbf{a}, Simulated component of the electric field of the even mode oriented in direction of epitaxial growth. \textbf{b}, Simulated component of the electric field of the odd mode oriented in direction of epitaxial growth. \textbf{c}, Simulated even and odd mode indices as function of the width of the gap between the waveguides. \textbf{d}, Simulated even and odd mode indices as function of the waveguide width. \textbf{e}, Computed power coupling coefficient as function of the gap width for the waveguide width of $10$ $\upmu$m. \textbf{f}, Computed power coupling coefficient as function of the waveguide width for the gap width of $1$ $\upmu$m.}
     \label{fig:SI_coupler}
\end{figure}

\begin{figure}[t!]
    \centering
    \includegraphics[width=1.22\textwidth,center]{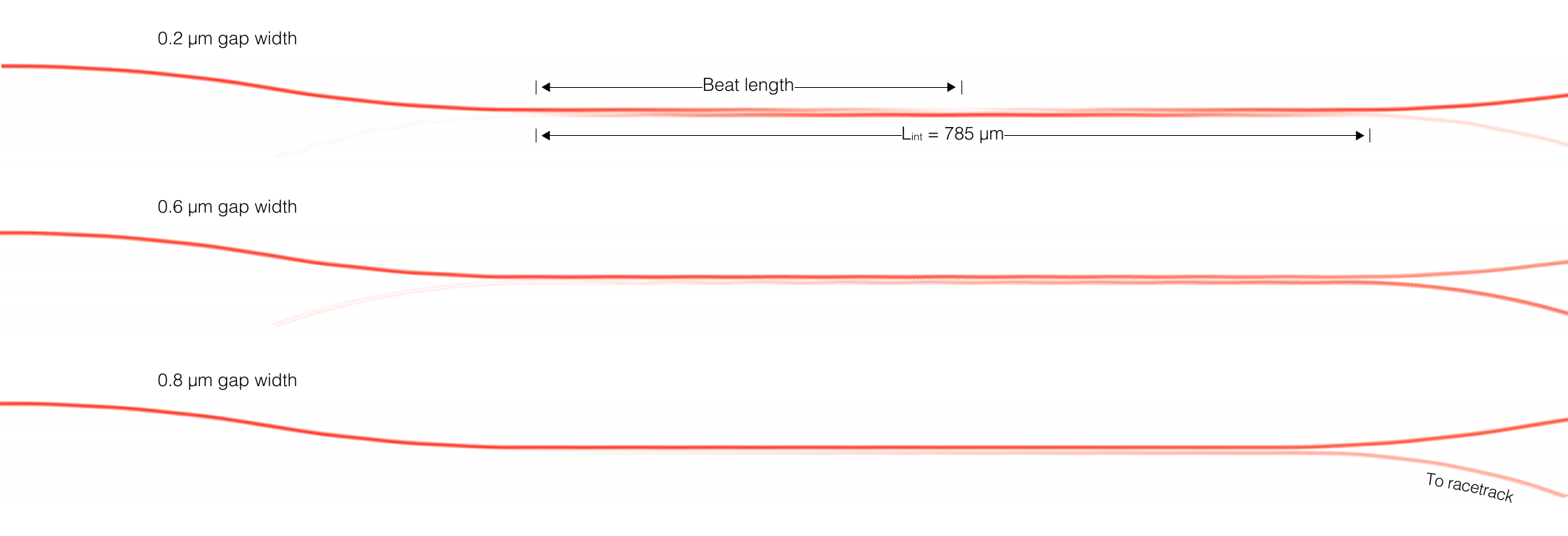}
    \centering
    \caption{2D COMSOL simulation of field intensity in a waveguide coupled to the racetrack for three different values of the gap width.}
     \label{fig:SI_2D_COMSOL}
\end{figure}

To optimize the coupler geometry, ultimately constrained by the resolution of the optical lithography, we first found the indices of the even ($n_e$) and odd ($n_o$) eigenmodes of the coupled waveguide structure using COMSOL (Fig.~\ref{fig:SI_coupler}\textbf{a}, \textbf{b}). Initially distinct values of $n_e$ and $n_o$ for closely spaced, strongly coupled waveguides, approach the same value as the waveguides get further apart (Fig.~\ref{fig:SI_coupler}\textbf{c}). The difference between $n_e$ and $n_o$ also gets more pronounced the narrower the waveguides are (Fig.~\ref{fig:SI_coupler}\textbf{d}). The contrast between the even and odd mode indices $k = n_e-n_o$ and the length of the interaction region $L_{\mathrm{int}}$ then define the power coupling, for the wavelength $\lambda = 7.9$ $\upmu$m, according to
\begin{equation}
    \left|\kappa\right|^2 = \frac{1}{1 + (\delta / k)^2} \sin^2{\left(\frac{\pi \sqrt{k^2 + \delta ^2}}{\lambda}L_{\text{int}}\right)},
    \label{eqn:coupling}
\end{equation}
where $\delta = n_{\mathrm{RT}} - n_{\mathrm{WG}}$ is the mode index mismatch in the coupled waveguides, that is set to zero in the simulation, but in the experiment may increase as we bias the \acrshort{wg} and \acrshort{rt} independently. Coupling coefficient can thus be increased by decreasing the separation between the waveguides (Fig.~\ref{fig:SI_coupler}\textbf{e}), decreasing the waveguide width (Fig.~\ref{fig:SI_coupler}\textbf{f}), and by increasing the length of the interaction region (Fig.~\ref{fig:SI_coupler}\textbf{e}, \textbf{f}). The system will become overcoupled for too large $L_{\mathrm{int}}$ or for very narrowly spaced waveguides (Fig.~\ref{fig:SI_2D_COMSOL}). 

\section{Circular ring QCLs with active couplers}
\begin{figure}[t!]
    \centering
    \includegraphics[width=1.0\textwidth,center]{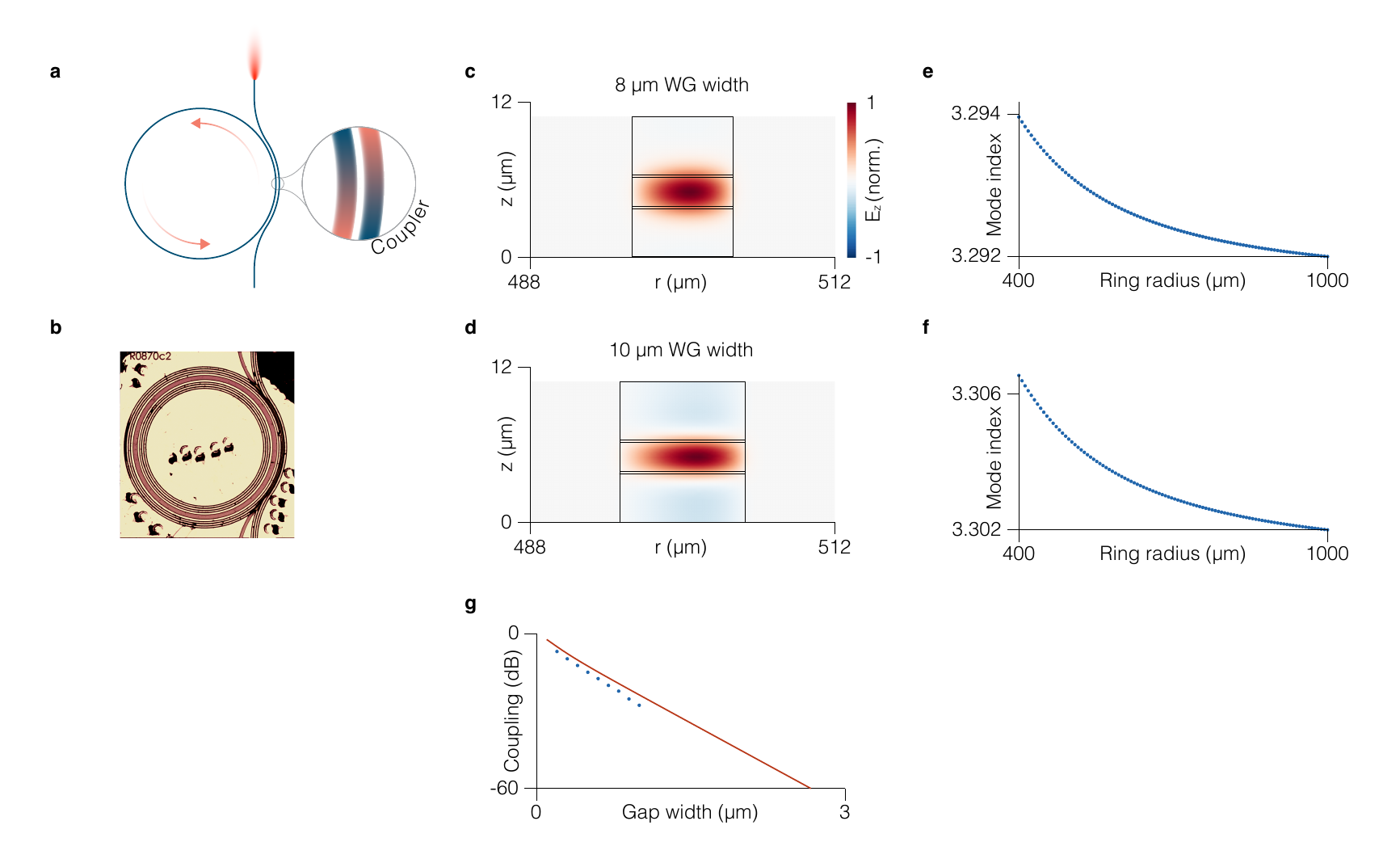}
    \centering
    \caption{\textbf{a}, Circular ring \acrshort{qcl} with a bent waveguide coupler implemented as a segment of a ring of a larger radius, concentric with the ring \acrshort{qcl}. \textbf{b}, Optical micrograph of a fabricated circular ring \acrshort{qcl} with a waveguide coupler. \textbf{c}, Simulated z-component (in cylindrical coordinates) of the electric field of the fundamental mode of the $8$ $\upmu$m wide waveguide of a circular ring with a radius of $500$ $\upmu$m. \textbf{d}, Same as in \textbf{c}, but for a waveguide width of $10$ $\upmu$m. The mode gets noticeably pushed to the outer boundary of the ring waveguide. \textbf{e}, Simulated mode index in a $8$ $\upmu$m wide circular ring waveguide as a function of the ring radius. \textbf{f}, same as in \textbf{e}, but for a waveguide width of $10$ $\upmu$m. \textbf{g}, Computed (solid line) and simulated (dots) power coupling coefficient for a circular ring as a function of gap width.}
     \label{fig:SI_circular_ring}
\end{figure}

Here we show that circular ring \acrshort{qcl}s can as well support an integration of a directional coupler. Here the coupler is implemented as a segment of a ring of a larger radius, concentric with the ring laser cavity (Fig.~\ref{fig:SI_circular_ring}\textbf{a}). We fabricated and tested such devices (Fig.~\ref{fig:SI_circular_ring}\textbf{b}), which behave similarly to \acrshort{rt} \acrshort{qcl}s of the main text. In this geometry, as opposed to the coupling section with two straight waveguide segments, an additional reduction in coupling may come from the mode index mismatch in the two waveguides with different radii of curvature. The lower is the radius of curvature the more the mode gets pushed away from the center of the waveguide towards the outer rim of the ring, which results in higher mode index (Fig.~\ref{fig:SI_circular_ring}\textbf{c}, \textbf{e}). This effect gets more pronounced for wider waveguides (Fig.~\ref{fig:SI_circular_ring}\textbf{d}, \textbf{f}). Similarly to the straight coupler geometry, the power coupling between two curved waveguides coefficient decreases with increasing gap width (Fig.~\ref{fig:SI_circular_ring}\textbf{g}). 

\section{Power output of a racetrack QCL with an active coupler}

\begin{figure}[t!]
    \centering
    \includegraphics[clip=true,width=1\textwidth]{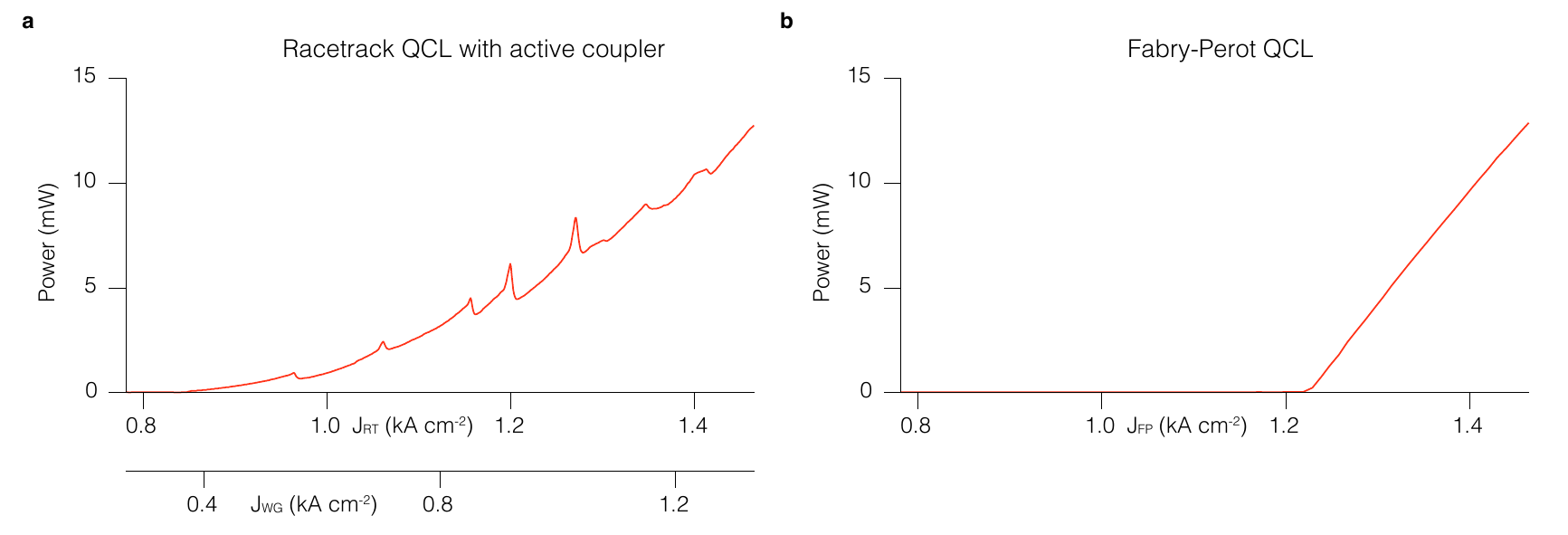}
    \caption{\textbf{a}, Output power from the front waveguide facet as both \acrshort{rt} and \acrshort{wg} currents are swept simultaneously. \textbf{b}, Output power of a Fabry-Perot QCL fabricated on the same wafer as the racetrack \acrshort{qcl} in \textbf{a}.}
     \label{fig:SI_LI}
\end{figure}

As described in the main text, \acrshort{rt} \acrshort{qcl}s with directional couplers have output optical power levels comparable to Fabry-Perot \acrshort{qcl}s. An attractive feature of the active \acrshort{wg} coupler is that it can additionally serve as an amplifier for the light coupled out from the \acrshort{rt}. When increasing the current of both \acrshort{wg} and \acrshort{rt} simultaneously, we observe an exponential increase in the \acrshort{rt} power output, as the \acrshort{wg} gain grows (Fig.~\ref{fig:SI_LI}\textbf{a}). The resonances in the power output are due to the fact that \acrshort{wg} itself is a Fabry-Perot resonator with end facet reflectivities of $0.3$. If necessary, the reflectivity can be reduced by the deposition of a quarter wave thick layer of a dielectric, such as Si$_3$N$_4$. The output power of a \acrshort{rt} \acrshort{qcl} is as large as that of a Fabry-Perot \acrshort{qcl} coming from the same fabrication batch (Fig.~\ref{fig:SI_LI}\textbf{b}).

\begin{figure}[t!]
    \centering
    \includegraphics[clip=true,width=0.5\textwidth]{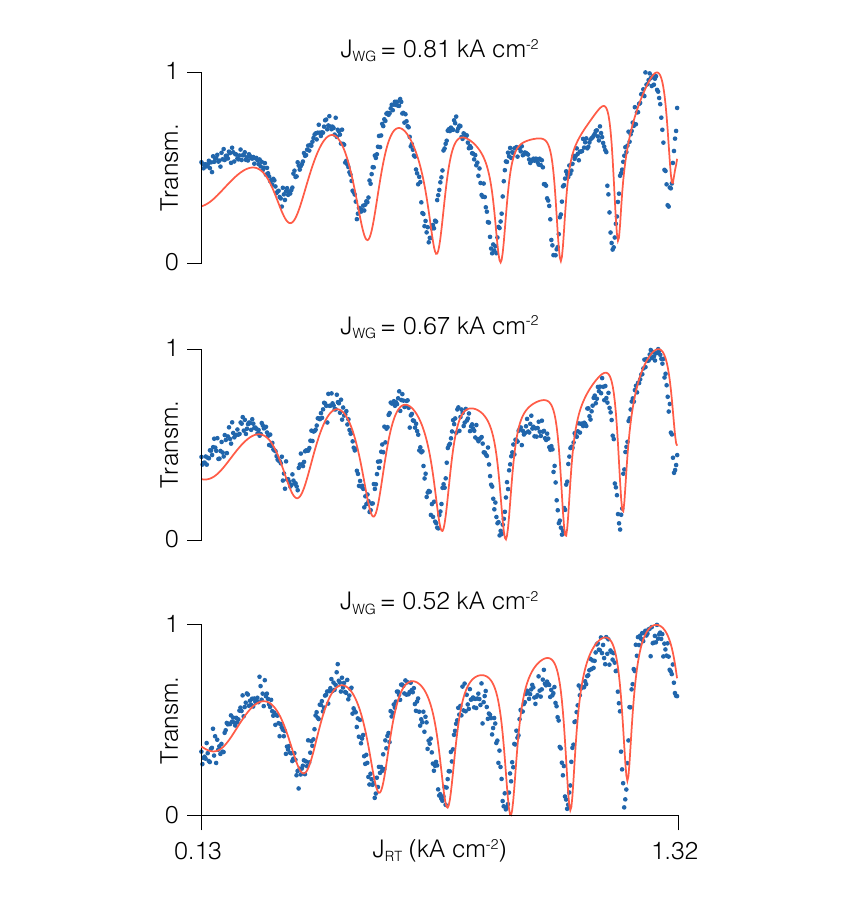}
    \caption{Optimized fitting (red) of Eq.~\ref{eqn:transmission} on three experimentally-obtained transmission spectra (blue), corresponding to $J_{\text{WG}} = 0.524 \text{ kA/cm}^2$, $J_{\text{WG}} = 0.668 \text{ kA/cm}^2$, $J_{\text{WG}} = 0.813 \text{ kA/cm}^2$.}
     \label{fig:SI_Fits}
\end{figure}
\section{Cavity transmission model and fitting}

We model and fit the RT-WG system below threshold using the transfer matrix method in order to extract various parameters of the system, notably the refractive index, coupling coefficient, and index crosstalk between the \acrshort{wg} and \acrshort{rt}.  The total transfer matrix, $M$, is a $2\times2$ matrix that relates the counter-propagating input field components, $a_\text{in}$ and $b_\text{in}$, to the output field components $a_\text{out}$ and $b_\text{out}$ via
\begin{align}
\begin{pmatrix} a_\text{out} \\ b_\text{out} \end{pmatrix} 
= M \begin{pmatrix} a_\text{in} \\ b_\text{in} \end{pmatrix}
\label{eqn:TMMtot}
\end{align}
$M$, is the product of all substituent transfer matrices that represent individual interactions in the system, $
M = M_{\text{F}}M_{\text{WG}}M_{\text{RT}}M_{\text{WG}}M_{\text{F}} $.  Here, $M_{\text{F}}$ represents interface scattering from either air to WG or WG to air, $M_{\text{WG}}$ represents propagation along half of the WG, and $M_{\text{RT}}$ represents coupling to and propagation of the mode in the RT resonator.

Assuming propagation along the $+x$ direction using the time convention of $\exp{\left\{i(\omega t - \beta x)\right\}}$, where $\omega$ is the angular frequency of the injected field and $\beta$ is the propagation constant, the individual transfer matrices are given by:
\begin{flalign}
\hspace{5cm} &M_{\text{F}} = \frac{1}{2 n_2} \begin{pmatrix} n_2 + n_1 &  n_2 - n_1 \\ n_2 - n_1 &  n_2 + n_1 \end{pmatrix},&
\label{eqn:facetM}
\end{flalign}
\begin{flalign}
\hspace{5cm} &M_{\text{WG}} = \begin{pmatrix} \exp{\left(-i \beta_{\text{WG}} L_{\text{WG}}/2\right)} &  0 \\ 0 &  \exp{\left(i \beta_{\text{WG}} L_{\text{WG}}/2\right)} \end{pmatrix},&
\label{eqn:wgM}
\end{flalign}
\begin{flalign}
\hspace{5cm} &M_{\text{RT}} = \begin{pmatrix}t_{\text{RT}} &  0 \\ 0 &  1/t_{\text{RT}} \end{pmatrix}, \\
&t_{\text{RT}} = \frac{\sqrt{1 - \kappa ^ 2} - \exp{\left(-i \beta_{\text{RT}}C_{\text{RT}}\right)}}{1 - \sqrt{1 - \kappa ^ 2}\exp{\left(-i \beta_{\text{RT}}C_{\text{RT}}\right)}}.&
\label{eqn:rtM}
\end{flalign}
In Eq.~\ref{eqn:facetM}, $n_1$ and $n_2$ are the (generally) complex mode indices of the two media on the opposite sides of the interface (air and the waveguide).  In Eq.~\ref{eqn:wgM}, $\beta_{\text{WG}} = \frac{2\pi n_{\text{WG}}}{\lambda}$ is the propagation constant of the WG, where $\lambda$ is wavelength of the injected field, and $L_{\text{WG}}$ is the length of the WG. In Eq.~\ref{eqn:rtM}, $\kappa$ is the amplitude coupling coefficient, that can be computed from Eq.~\ref{eqn:coupling}, $\beta_{\text{RT}}$ is the propagation constant of the RT, and $C_\text{RT}$ is the circumference of the RT.  Assuming no initial counter-propagation at the output ($b_{\text{out}} = 0$), the complex transmission coefficient is given by:
\begin{align}
\frac{a_{\text{out}}}{a_{\text{in}}} = \frac{1}{M(2, 2)}\left(M(1, 1)M(2, 2) - M(1, 2)M(2, 1)\right).
\label{eqn:transmission}
\end{align}
Eq.~\ref{eqn:transmission} is used to fit the experimental transmission of the RT-WG system shown in the main text.  

In order to accurately reproduce the experimental data, we assume the refractive index of either the WG or RT changes as a function of current density, $J$, to an arbitrary power:
\begin{align}
&n_{\text{RT}}(J_\text{RT}) = n_{\text{RT}_0} + \Delta n_{\text{RT}} J_\text{RT}^\alpha,\\
&n_{\text{WG}}(J_\text{WG}) = n_{\text{WG}_0} + \Delta n_{\text{WG}} J_\text{WG}^\alpha,
\label{eqn:nsweep}
\end{align}
where $n_{\mathrm{RT,WG}_0}$ is the base refractive index and $\Delta n_{\mathrm{RT,WG}}$ is how much the index changes per unit $J^{\alpha}$.  We also model crosstalk between the RT and WG, i.e. how heating of either the RT or WG due to biasing affects the refractive index of the other section.  It is accounted for by additional contribution, $\Delta n_{\text{cross}}$, to the refractive index of either the RT or WG, which is directly proportional the current density of the other section:
\begin{align}
&n_{\text{RT}}(J_\text{RT}, J_\text{WG}) = n_{\text{RT}}(J_\text{RT}) + \Delta n_{\text{cross}} J_\text{WG}\\
&n_{\text{WG}}(J_\text{WG}, J_\text{RT}) = n_{\text{WG}}(J_\text{WG}) + \Delta n_{\text{cross}} J_\text{RT}.
\label{eqn:nsweep}
\end{align}

We select three experimentally-measured transmission spectra (corresponding to $J_{\text{WG}} = 0.524 \text{ kA/cm}^2$, $J_{\text{WG}} = 0.668 \text{ kA/cm}^2$, $J_{\text{WG}} = 0.813 \text{ kA/cm}^2$) to preform least squares fitting using Eq.~\ref{eqn:transmission}, as shown in Fig.~\ref{fig:SI_Fits}.

\bibliography{racetrack_references}